\documentclass{article}
\usepackage{amstext}
\usepackage{amsmath}
\usepackage{graphicx}
\usepackage{mathrsfs}
\usepackage{indentfirst}
\usepackage{amssymb}
\usepackage{color}
\usepackage{cite}

\title{Single production of dark photon at the LHC}
\author{Zhi-Cheng Liu, Chong-Xing Yue\thanks{E-mail:cxyue@lnnu.edu.cn},  and  Yu-Chen Guo\\
{\small Department of Physics, Liaoning  Normal University, Dalian
116029, P. R. China}
}
\date{\today}

\begin{document}
\maketitle
\begin{abstract}
Many extensions of the SM contain hidden sectors which are uncharged under the SM gauge group. The simplest model with an additional $U(1)$ group predicts the existence of  the new gauge boson, dark photon $Z_D$. In this work we consider single production of dark photon $Z_D$ at the LHC and study its discovery potential  via the process $pp\rightarrow Z_D\gamma\rightarrow ll\gamma$. After simulating the signal as well as the relevant backgrounds, the numerical results show that the signal of $Z_D$ might be detected at the 14TeV LHC.
\end{abstract}

  \section{Introduction}
Although most of the experimental measurements are in good agreement with the standard model (SM) predictions, there are still some unexplained  discrepancies and theoretical issues that the SM can not solve\cite{SM}. So the SM is generally regarded as an effective realization of an underlying theory to be  yet discovered. The small but  non-vanishing neutrino masses, the hierarchy and naturalness problems provide a strong motivation for contemplating new physics (NP) beyond the SM at TeV scale, which would be in an energy range accessible at the LHC.

  Many extensions of the SM contain hidden sectors. These corresponding hidden particles are uncharged under the SM gauge group, and can be natural candidates of dark matter. They can interact with SM sectors indirectly by some gauge invariant mixing terms. The SM has two neutral operators, the hypercharge field-strength $B_{\mu\nu}$ and the Higgs modulus squared  $\left|H\right|^2$. If the hidden sectors have  $U(1)_D$  gauge field  $Z_D$, and  scalar field $S$, they could couple to the former SM neutral operators in a  renormalizable way: $Z_{D\mu\nu}B^{\mu\nu}$ and $\left|H\right|^2\left|S\right|^2$\cite{2U1}.  The other  renormalizable operators involve $G_{i\mu\nu}\tilde{G}^{i\mu\nu}$, $\partial_{\mu}a\bar{\psi}\gamma^{\mu}\gamma^5\psi$ and $LHN$\cite{1311.0029}, here $G_{i\mu\nu}$ is the gluon field strength tensor, $a$ is SM singlet scalar, $L$ is SM lepton and $N$ is sterile neutrino. They can explain the generation of strong CP violation and neutrino mass. These coupling operators give us an opportunity to probe the hidden world. In this work we consider the simplest model which only contains a  $U(1)_D$  gauge field as well as  a scalar field $S$ and can be an effective theory of many particular new physics models. There are many theoretical and experimental efforts on this kind of models \cite{2U1,1311.0029,0803.1243,accel-1,accel-2,accel-3,0801.3456,1405.5196,1405.7691,1408.1075,1312.4992,1603.01377, 1412.0018,0810.0714,1505.07645,0903.3941,exp,1209.6083}. In this work, we study  single production of  $Z_D$ at the 14TeV LHC and give the relevant production cross sections. We further discuss the signature of the process $pp\rightarrow Z_D\gamma\rightarrow ll\gamma$ and the SM  backgrounds. Our numerical results show that the signal of $Z_D$ might be detected at the 14TeV LHC.

  The rest of this paper is organized as follows. In section 2, we review the elementary feature of the simple dark photon model, summarize the constraints on its free parameters and possible decay channels  of the  dark photon $Z_D$. The relevant calculation formula and numerical results are given in section 3. Conclusions are presented in section 4.

  \section{Simple dark photon model} The simple dark photon model\cite{1412.0018} is based on the gauge group $SU(3)_C\times SU(2)_L\times SU(2)_Y\times U(1)_D$. A dark Higgs field $S$ with non-zero vacuum expectation value(VEV) is introduced to break the dark $U(1)_D$ symmetry. The relevant terms in the Lagrangian, which are related to our calculation, are
  \begin{align}
  \mathcal{L}_{gauge}&\supset-\frac{1}{4}\hat{B}_{\mu\nu}\hat{B}^{\mu\nu}-\frac{1}{4}\hat{Z}_{D\mu\nu}\hat{Z}_{D}^{\mu\nu}+\frac{1}{2}\frac{\epsilon}{c_w^2}
  \hat{Z}_{D\mu\nu}\hat{B}^{\mu\nu},\\
  \mathcal{L}_{scalar}&\supset-\mu^{2}\left|H\right|^2+\lambda\left|H\right|^2-\mu_S^2\left|S\right|^2+\lambda_S\left|S\right|^4+\kappa\left|H\right|^2\left|S\right|^2,
  \end{align}
where the hatted fields are gauge eigenstates, $\hat{B}_{\mu\nu}=\partial_{\mu}\hat{B}_{\nu}-
  \partial_{\nu}\hat{B}_{\mu}$ and $\hat{Z}_{D\mu\nu}=\partial_{\mu}\hat{Z}_{D\nu}-\partial_{\nu}\hat{Z}_{D\mu}$ represent the hypercharge $U(1)_Y$ and dark $U(1)_D$ field strengths, respectively, $c_w$ is the cosine of Weinberg angle $\theta_w$,  $\epsilon$ is the kinetic mixing parameter, $H$ is the SM Higgs doublet, $S$ is the SM-singlet dark Higgs, and $\kappa$ is the Higgs mixing parameter.
  Electroweak symmetry is broken by $\left \langle H \right \rangle=\left(0,v/\sqrt{2}\right)$, where $v\approx  246 $ GeV. The dark $U(1)_D$ symmetry is broken by $\left \langle S \right \rangle=v_{S}/\sqrt{2}$. The mass eigenstates of gauge boson can be obtained by rotating the three neutral components of the gauge fields $\hat{Z}_D$, $\hat{B}$ and $W_3$,
   
\begin{equation}
\setlength{\arraycolsep}{3.5pt}
 \begin{pmatrix}
  Z_D\\A\\Z
  \end{pmatrix}
=\left(
	\begin{array}{ccc}
      c_{\alpha} & 0 & -s_{\alpha}\\
	  0 & 1 & 0 \\
	  s_{\alpha} & 0 & c_{\alpha}	
	\end{array}
\right)
\left(
	\begin{array}{ccc}
	1 & 0 & 0\\
 	0 & c_w & s_w \\
 	0 & -s_w & c_w
	\end{array}
\right)
\left(
	\begin{array}{ccc}
	\sqrt{1-\frac{\epsilon^2}{c_w^2}} & 0 & 0 \\
 0 & 1 & 0 \\
 \frac{\epsilon}{c_w} & 0 & 1	
	\end{array}
\right)
\begin{pmatrix}
  \hat{Z}_D\\\hat{B}\\W_3
  \end{pmatrix},
\end{equation}
the $Z_D$,$A$ and $Z$ are mass eigenstates of the gauge fields.
The first and second matrices are used to diagonalize the mass terms and the third matrix is used to canonically normalize the kinetic terms. The $s_{\alpha}$ and $c_{\alpha}$ are sine and cosine of the new mixing angle $\theta_\alpha$ given by
 \begin{equation}
 \tan\left(2\theta_\alpha\right)=\frac{-2s_w\eta}{1-s_w^2\eta^2-\Delta^2},
 \end{equation}
 where  $\eta=\epsilon/\sqrt{c_w^2-\epsilon^2}$, $\Delta^2=M_{D0}^2/M_{Z0}^{2}$,  $M_{Z0}$ and $M_{D0}$ are the masses of the SM boson $Z$  and the dark boson before mixing. After the above rotating, the SM photon field $A$ remains massless, the mass eigenvalues of the SM boson $Z$  and dark photon $Z_D$ are  
\begin{equation}
M_{Z,Z_D}^2=\frac{M_{Z0}^2}{2}\left\{1+\Delta^2+\eta^2 s_w^2\pm Sign(1-\Delta^2)\sqrt{(1+\Delta^2+\eta^2s_w^2)^2-4\Delta^2}\right\}
.
\end{equation}
The interaction between the dark photon  $Z_D$  and the SM particles can be obtained by an expansion of the Lagrangian in terms of mass eigenstates. For example, the interaction between the dark photon  $Z_D$  and the SM fermions is
 \begin{align}
\mathcal{L}_{Z_D\bar{f}f}&=g_{Z_D\bar{f}f}Z_{D\mu}\bar{f} \gamma^\mu f,
\\
g_{Z_D\bar{f}f}&= \frac{g}{c_w}\left\{-s_{\alpha} (T_{3L} c_w^2-Ys_w^2)+Y\eta c_{\alpha} s_w \right\}.
\end{align}
where $f$ denotes the left- or right- fermion, $T_{3L}$ and $Y$ are the third component of weak isospin and hypercharge of the fermion $f$, respectively.
More Feynman rules can be found in Ref.\cite{0801.3456}.
 
 The relevant parameters related to our calculation are $\epsilon$ and $M_{Z_D}$. We assume that the dark Higgs is very heavy and the scalar mixing parameter $\kappa$ is sufficiently small, so they do not affect  the dark photon phenonenology. 
  Dark photon with $M_{Z_D}\sim$ MeV may explain the $\sim 2\sigma$ excess of dark radiation in the universe\cite{2sigma}. The related experimental probes  include precision QED measurements, rare meson decays, supernova cooling, collider experiments and beam dumps\cite{0903.3941,exp,1209.6083}. Dark photon with mass $\sim$GeV   could explain the $\sim 3.6\sigma$ deviation between the observed and SM values of the muon anomalous magnetic moment\cite{gu2}. Moreover, it might explain possible terrestrial and cosmic ray dark matter anomalies\cite{DM}. The related experimental probes of dark photon in this mass scale include new accelerator based experiments\cite{accel-1}, especially with new beam dump and fixed target experiments exploiting high intensity electron \cite{accel-2} and proton beams\cite{accel-3}.The Babar Collaboration has search for dark photon with mass range 0.02-10.2 GeV\cite{Babar}. In the case of $M_{Z_D}$ above 10 GeV, the LHC and future high energy collider offer opportunities to probe dark photon because of their high energy and luminosity.  Refs.\cite{1701.08614} has estimated the sensitivity of $Z_D$ at the CEPC and FCC-ee through the process  $e^+e^-\rightarrow Z_D\gamma \rightarrow \mu^+\mu^-\gamma$.  Exotic decays of Higgs $H\rightarrow Z_DV$ (with $V=Z,Z_D,\gamma$) were explored in Refs.\cite{0801.3456,1405.5196,1405.7691,1408.1075,1312.4992,1603.01377, 1412.0018}. ATLAS collaboration has searched for  Higgs   decay channels, $H\rightarrow ZZ_D\rightarrow 4l$ and $H\rightarrow Z_DZ_D\rightarrow 4l$, but no significant signal is observed\cite{1505.07645}. Refs.\cite{1408.1075,1412.0018} have also estimated the sensitivity of $Z_D$ at the LHC through Drell-Yan production process. 
 
 The free parameter $\epsilon$ denotes the strength of kinetic mixing. Since the kinetic mixing term is dimension-4 operator and unsuppressed by any high mass scale,  $\epsilon$  theoretically is not required to be small. In particular models, the parameter $\epsilon$ can be calculated by perturbative or non-perturbative ways\cite{1311.0029}.
For example, when the mixing interaction is generated at a high scale in a Grand Unified Theory(GUT) or string theory, one can obtain $\epsilon$ in a wide range of $10^{-17}$-$10^{-2}$\cite{2U1,0903.3941,0810.0714,EPSILON}.
 On the experimental side, when the mass $M_{Z_D}$ is larger than  10 GeV, the most strict constraints on $\epsilon$ come from electroweak precision observables. As shown in Refs.\cite{1408.1075,1412.0018}, the upper limit of $\epsilon$ is about 2-3$\times 10^{-2}$.  ATLAS collaboration has searched exotic Higgs decays  $H\rightarrow Z_DZ_D$ and $H\rightarrow ZZ_D$. Upper bound on $\epsilon$ is set in the range 4-7$\times 10^{-2}$ for 10 GeV$ <M_{Z_D}< $ 35  GeV\cite{1505.07645}. At higher masses, the LHC searches for high-mass dilepton resonances can be used to constrain the  parameters. In Fig.1 we show the 95\% C.L. excluded region in $\epsilon - M_{Z_D}$ plane by ATLAS\cite{1607.03669}. To obtain this constraints, we have used MADGRAPH5\cite{MG5} to generate simulated events. The 13TeV dilepton analysis was implemented into the CheckMATE1.2.2 framework\cite{1503.01123}. 
 For latter collider simulation, we consider a benchmark parameter point $(\epsilon=0.03, M_{Z_D}=400\ \text{GeV})$ as shown in Fig.1.
\begin{figure}
\begin{center}
\includegraphics[scale=0.8]{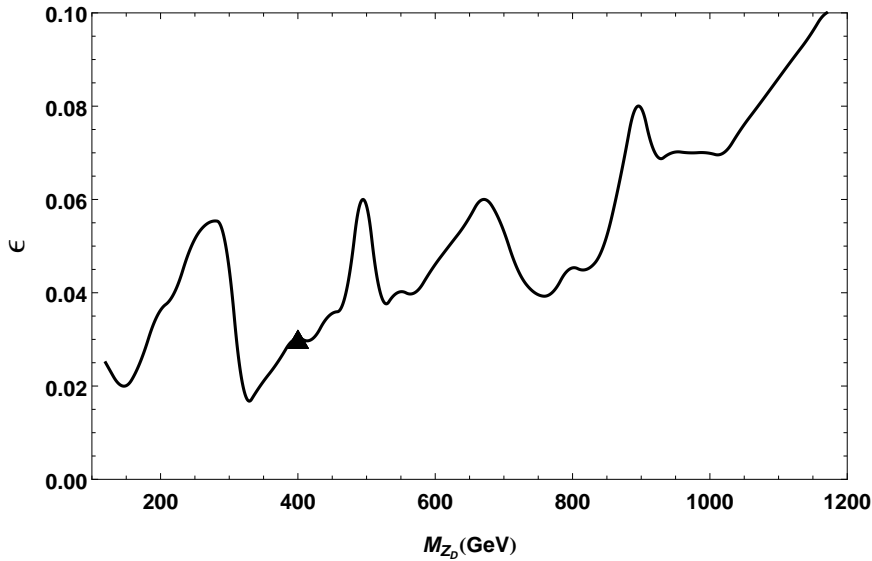}
\caption{The 95\% C.L. excluded region in $\epsilon - M_{Z_D}$ plane from ATLAS data. The benchmark parmeter point is also indicated.}
\end{center}
\end{figure}
 
 Dark photon $Z_D$ with high mass can decay to a pair of the SM particles. When $\epsilon\ll1$, the  coupling  between dark photon $Z_D$ and the SM particles is approximately proportional to the tiny parameter $\epsilon$, this feature leads to a rather small width of $Z_D$. But the branching ratios (BRs)  are independent of $\epsilon$. In Fig.2 we show the branching ratios of $Z_D$  as a function of $M_{Z_D}$ at leading order, where $l$ refers to  $e$ or $\mu $, $d$ the down type quark and $u$ refers to the up type quark except the top quark. The BRs are approximately constants when $M_{Z_D}$ is above 350 GeV.

\begin{figure}
\begin{center}
\includegraphics[scale=0.8]{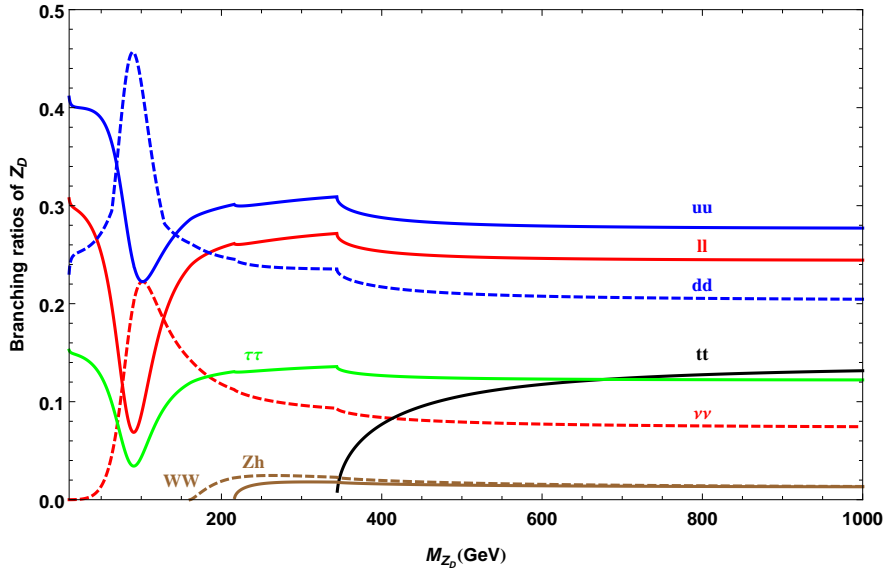}
\caption{Branching ratios (BRs) of dark photon $Z_D$ at the tree level. }
\end{center}
\end{figure}

 \section{Single production of dark photon  at  the LHC}

	At the LHC  dark  photon can be produced  via many processes, such as Drell-Yan production process, production in association with a vector boson $Z_D+\gamma/Z/W$ or with jets $Z_D+$jets, and exotic decays of Higgs $pp\rightarrow H\rightarrow ZZ_D$, or $pp\rightarrow H\rightarrow Z_DZ_D$. The  process $H\rightarrow Z_DZ_D$ depends on the parameter  $\kappa $ and $M_{Z_D}$, the other processes depend on the parameter $\epsilon$ and $M_{Z_D}$. Exotic decays of Higgs  have been explored in Refs.\cite{0801.3456,1405.5196,1405.7691,1408.1075,1312.4992,1603.01377, 1412.0018}, so we consider $Z_D$ production in other processes.
	
	 The effective cross section $\sigma\left(s\right)$ can be evaluated from $\hat{\sigma}\left(\hat{s}\right)$ by convoluting with the parton distribution functions $f_{q_1/p}\left(x_1\right)$ and $f_{q_2/p}\left(x_2\right)$,
 \begin{equation}
 \sigma\left(s\right)=\int^1_{x_{min}}dx_1\int^1_{x_{min}/x_1}dx_2f_{q_1/p}\left(x_1\right)f_{q_2/p}\left(x_2\right)
 \hat{\sigma}\left(\hat{s}\right),\end{equation}
where $\hat{\sigma}\left(\hat{s}\right)$ refers to the partonic cross section of subprocesses $q\bar{q}\rightarrow Z_D+\gamma/Z/W/$jet, $\hat{s}=x_1x_2s$ is the effective center-of-mass energy squared for the partonic process, and $x_{min}=M^2/s$, $M$ is sum of mass of the final particles.
For parton distribution functions $f_{q_1/p}\left(x_1\right)$ and $f_{q_2/p}\left(x_2\right)$, we used the forms given by the leading order parton distribution function CTEQ6L1\cite{CTEQ}. The cross sections versus $M_{Z_D}$ are showed in Fig.3. Here we have used MADGRAPH5\cite{MG5} to calculate cross sections and imposed minimum transverse momentum cuts on photons and jets,  $p_T^{\gamma}>10 $ GeV and $p_T^j>20 $ GeV.
\begin{figure}
\begin{center}
\includegraphics[scale=0.8]{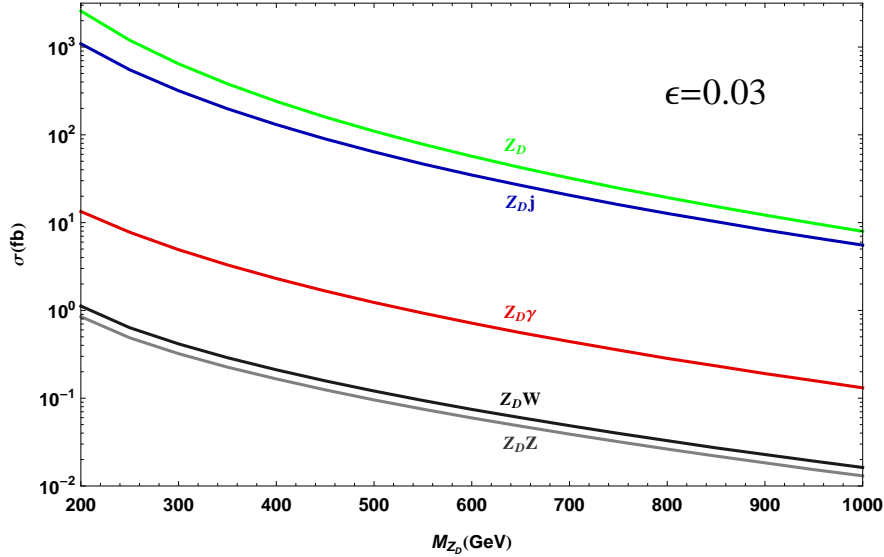}
\caption{The production cross sections of the dark photon $Z_D$ with $\epsilon=0.03$ at the 14TeV LHC. }
\end{center}
\end{figure}

According to Fig.2, we can see that  $Z_D\rightarrow q\bar{q}$ is the main decay channel of $Z_D$ with mass range $M_{Z_D}>100$ GeV. But it is hard to be detected at the LHC owing to the huge QCD backgrounds. Unlike the SM boson $Z$ , the leptonic branching ratio of $Z_D$ is  sizable, about 25\%. Furthermore, the leptonic channel has reduced backgrounds due to lepton energy and momentum resolution being better than jet energy resolution, so dilepton is the best search channel of $Z_D$.

 Among the  processes shown in Fig.3, the DY production process $pp\rightarrow Z_D\rightarrow ll$  is the best  discovery channel for $Z_D$. Refs.\cite{1408.1075,1412.0018} have estimated the potential sensitivity of $Z_D$ at the 14TeV LHC through this process. The process $Z_Dj$ is not suitable for detecting dark photon because of the overwhelming QCD backgrounds.

For $Z_DW$ pair production followed by $W\rightarrow l \nu_l$ and $Z_D\rightarrow ll$ decay. The signal events are characterized by three charged leptons and missing energy $3l+/\kern-0.65em E_T$. There are several major sources of SM backgrounds in $WZ_D$ trilepton events. The irreducible background is the  $WZ$ pair production. The reducible backgrounds involve, (1) $ZZ$, when one of the leptons from the $Z$ boson decay is not identified, (2) $W/Z+\gamma$, in which the photon produces  electrons via conversion, (3) $Z+$ jets, $WW$, $t\bar{t}$ and single-top-quark production when jets are misidentified as leptons.

The process $pp\rightarrow Z_DZ$ with four charged leptons in final state  has fewer events because of the low leptonic branching ratio of $Z$. The backgrounds involve, (1)irreducible background $ZZ$, (2)$Z+b\bar{b}$ and $t\bar{t}$, where the final states contain two isolated leptons and two b-quark jets producing secondary leptons, (3) $Z+$ jets and $WZ+$jets, where jets are misidentified as leptons.

For the process $pp\rightarrow Z_D\gamma\rightarrow ll\gamma$. Backgrounds of the $ll\gamma$ signal originate mainly from the following processes: (1) $pp\rightarrow Z\gamma$, (2) $pp\rightarrow Z+$jets, when one of the jets is misidentified as a photon, (3) $t\bar{t}\gamma$. The first one is irreducible and the last two are reducible.

 As a complementary to the DY production, we simulate the process $pp\rightarrow Z_D\gamma\rightarrow ll\gamma$ in detail due to its clear signature and large cross section.  In order to investigate the signature of $Z_D\gamma$ production at the LHC, we chose the 
 benchmark points $(\epsilon=0.03, M_{Z_D}=400\ \text{GeV})$  as shown in Fig.1.  For the following results we use the FeynRules\cite{FR} to implement the  simple dark photon  model\cite{1312.4992,1412.0018,0803.1243}, MADGRAPH5\cite{MG5} to generate signal and background events,  PYTHIA\cite{PYTHIA}  for parton shower and  hadronization and DELPHES\cite{DELPHES} as 
fast detector simulation with ATLAS detector card. Finally, MADANALYSIS5\cite{MA5} is applied for data analysis and plotting.
  The average efficiency of single photon is taken as 90\%. The identification efficiencies for leptons $e$ and $\mu$ are taken to be 90\% and 95\%, respectively. For $Z+$ jets events, we assume  a jet-faking-photon rate as 0.12\%. To aviod double counting between multiple jets and parton shower, the MLM matching method with merging scale $x_q$ = 15 GeV is applied.

  We pick up the events that include exactly one photon, one opposite-sign same-flavor lepton pair and one jet at the most, then impose the following  basic cuts,
\begin{equation}
p_T^{l,\gamma}>10 \ \text{GeV} ,\  \left | \eta^{l,\gamma} \right |<2.5,\ \Delta R_{ll}>0.4,\ \Delta R_{\gamma \gamma}>0.4,
\end{equation}
where $\Delta R=\sqrt{(\Delta\phi)^2+(\Delta\eta)^2}$ is the particle separation, $\Delta\phi$ is the rapidity gap and $\Delta\eta $ is the azimuthal angle gap between the particle pair.
\begin{figure}
\begin{center}
\includegraphics[scale=0.3]{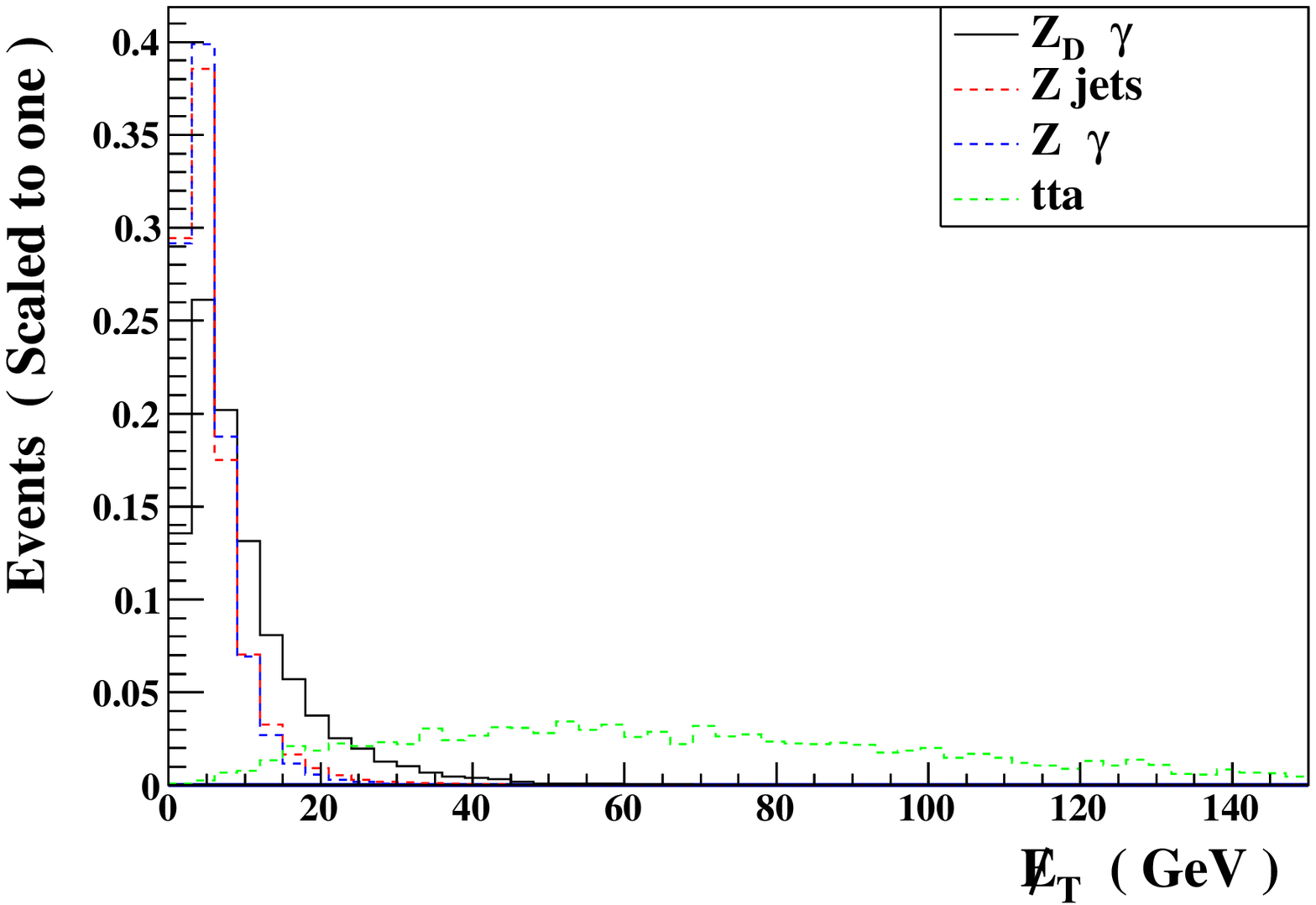}
\includegraphics[scale=0.3]{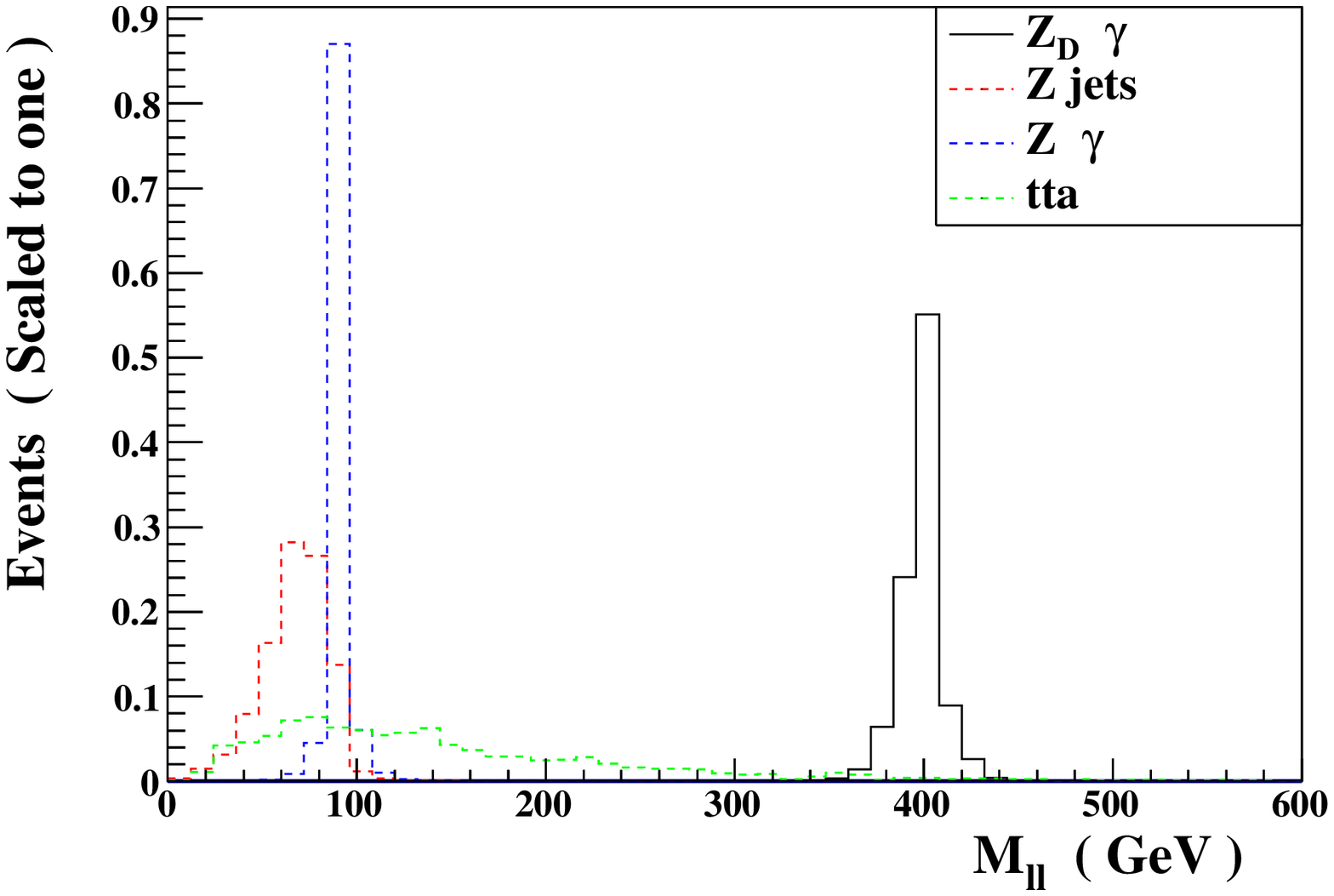}
\caption{Distributions of $/\kern-0.65em E_T$ (left) and $M_{ll}$ (right) for the signal and backgrounds after the basic cuts at the  14TeV LHC with $\mathcal{L}=300 fb^{-1}$.}
\end{center}
\end{figure}

\begin{figure}
\begin{center}
\includegraphics[scale=0.3]{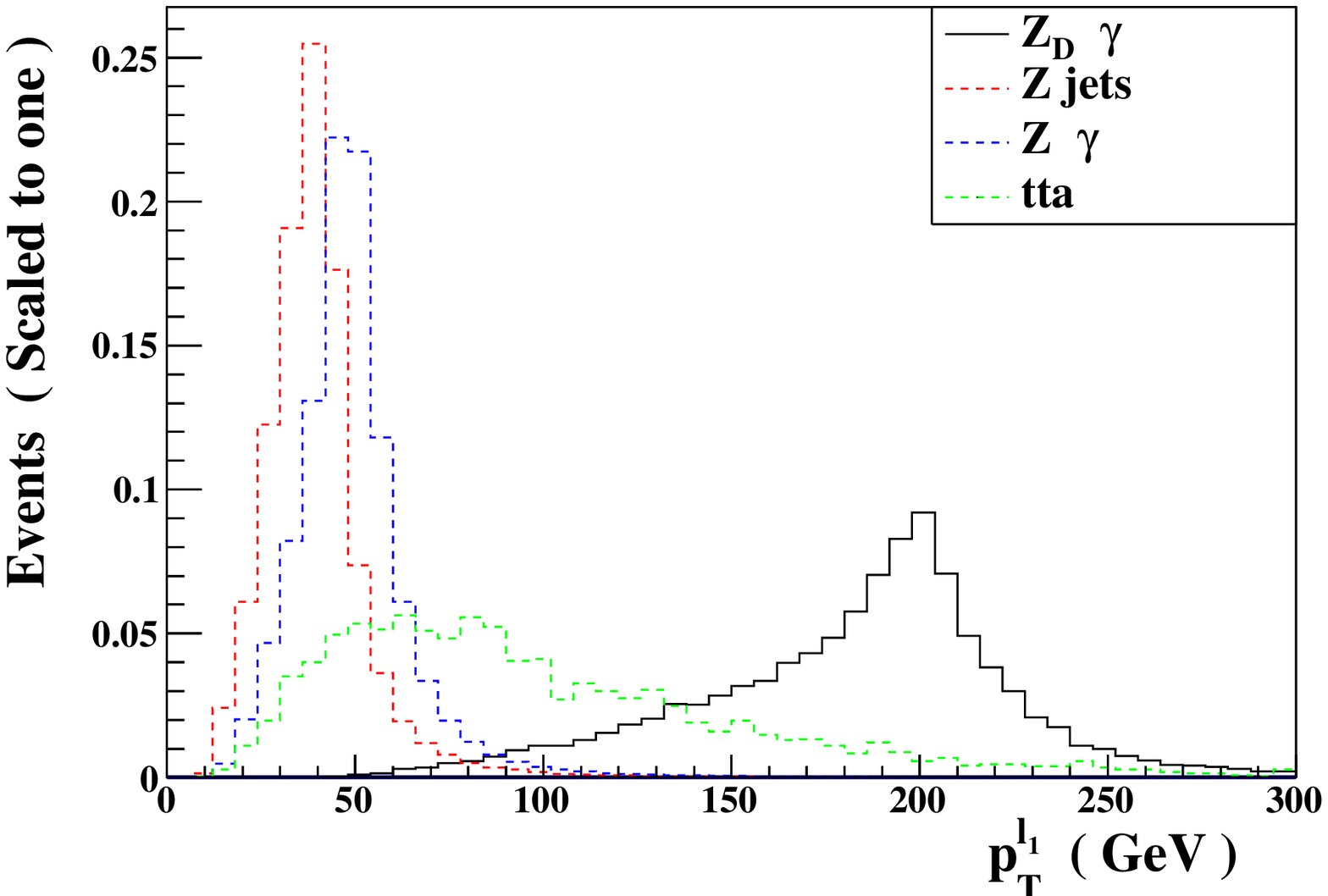}
\includegraphics[scale=0.3]{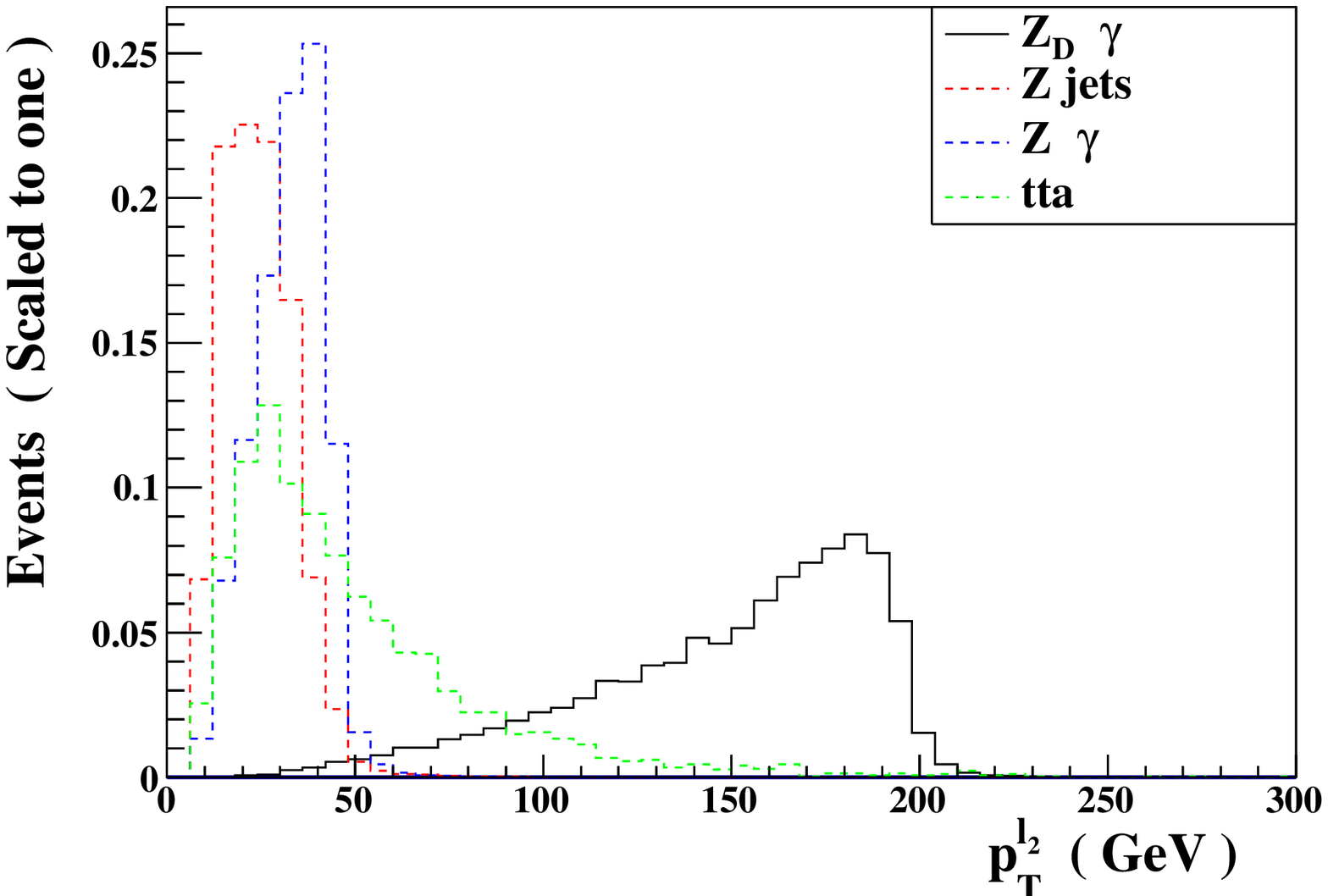}
\caption{Distributions of the leading lepton transverse momentum $p_T^{l_1}$(left) and the next-to-leading lepton transverse momentum $p_T^{l_2}$(right) for the signal and backgrounds after the basic cuts at the 14TeV LHC with $\mathcal{L}=300 fb^{-1}$.}
\end{center}
\end{figure}

\begin{table}[!hbp]
\begin{center}
\caption{The event numbers of the signal and backgrounds at the 14TeV LHC with $\mathcal{L}=300 fb^{-1}$.}
\begin{tabular}{c|c|c|c|c}

\hline
 &Signal $Z_D\gamma$&Bkg $Z\text{jets}$&Bkg $Z\gamma$&Bkg $tt\gamma$\\

\hline
Basic cuts                  &54.2   &2727082   &417183 &2058.5 \\
$\Delta R_{l\gamma}>0.7$,\ $/\kern-0.65em E_T<40\ \text{GeV}$  &53.4   &2719661  &416944 &461.1    \\
$p_T^{l_1}>70\ \text{GeV}$  &53.0  &82346  &30001 &301.9    \\
$p_T^{l_2}>50\ \text{GeV}$        &52.4   &14016  &2854.1 &181.9     \\
$\left |M_{ll}-M_{Z_D}\right |<20\ \text{GeV}$
                           &48.2  &147.2   &1.5  &3.9     \\

\hline
$S/\sqrt{S+B}$&\multicolumn{4}{|c}{3.4}\\
\hline

\end{tabular}
\end{center}
\end{table}
To further suppress the backgrounds, we shall apply several kinematical cuts. Firstly, the angular separation $\Delta R_{l\gamma}>0.7$ is  applied to reduce the contribution from final state photon radiation. In Fig.4 we show the distribution of the missing transverse momentum, only the $t\bar{t}\gamma$ events can be discriminated effectively. Since the contribution of the $t\bar{t}\gamma$ events is small, we choose very  loose missing transverse momentum cut $/\kern-0.65em E_T<40$ GeV.
Secondly, for the vaules of the parameters $\epsilon$ and $M_{Z_D}$ we use in this work, the SM boson $Z$  is lighter than the dark photon $Z_D$, so the final leptons of backgrounds are  softer than those of signal. In Fig.5 we show the distributions of the leading lepton transverse momentum $p_T^{l_1}$ and the next-to-leading lepton transverse momentum $p_T^{l_2}$. We can use this character to distinguish  the signal from backgrounds, through the following cuts,
\begin{equation}
p_T^{l_1}>70\ \text{GeV},\ p_T^{l_2}>50\ \text{GeV}.
\end{equation}
 Finally, we set a window on the dilepton invariant mass,
 \begin{align}
 \left |M_{ll}-M_{Z_D}\right |<20 \ \text{GeV}.
 \end{align}

 After all kinematical cuts are applied, the event numbers of signal and the backgrounds are summarized in Table I. We defined the statistical significance as $S/\sqrt{S+B}$ where $S$ and
$B$ denote the number of signal and background events, respectively. It can reach 3.4 at the 14TeV LHC with an integrated luminosity of 300 $fb^{-1}$. Fig.6 shows the expected 95\% CLs sensitivity region in $\epsilon - M_{Z_D}$ parameter space from $pp\rightarrow Z_D\gamma \rightarrow ll\gamma$ at the 14TeV LHC with $\mathcal{L}=100 fb^{-1}$ and $300 fb^{-1}$. The other processes offer weaker sensitivity. The $\epsilon$ of the processes $Z_DZ$ and $Z_DW$ is 3-9 times bigger than that of $Z_D\gamma$ at the same $M_{Z_D}$, so we do no display the sensitivity region of these processes. By comparing the Fig.1 and Fig.6, we can see that the Drell-Yan process of $pp\rightarrow Z_D\rightarrow ll$ is the most sensitive channel for the $Z_D$ searches. The process $pp\rightarrow Z_D \gamma$  can also be detected with high luminosity LHC data.

\begin{figure}
\begin{center}
\includegraphics[scale=0.8]{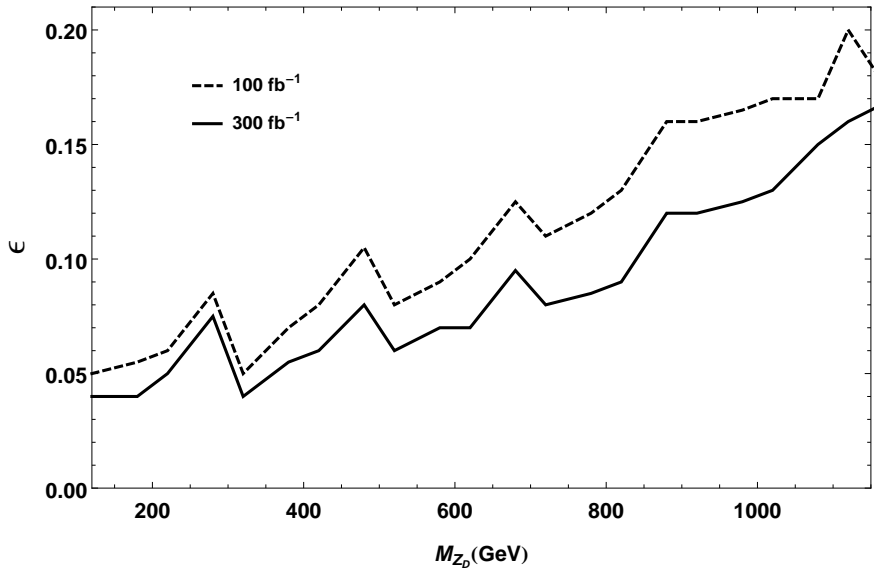}
\caption{Expected 95\% CLs sensitivity region in $\epsilon - M_{Z_D}$ parameter space from $pp\rightarrow Z_D\gamma \rightarrow ll\gamma$ at the 14TeV LHC with $\mathcal{L}=100 fb^{-1}$ and $300 fb^{-1}$.}
\end{center}
\end{figure}

\section{Conclusions}
	Many extensions of the SM contain hidden sectors which are uncharged under the SM gauge group. It is natural to suppose that it has some gauge interactions in the hidden sector.
The simplest model  contains a spontaneously broken dark  $U\left ( 1 \right )_{D}$ gauge symmetry which mediated by dark photon $Z_D$.  The new particles  can couple to the SM sectors by some mixing operators. This model can also be effective theory of many particular new physics models.
	
	 In this work we study single production of $Z_D$ at the 14TeV LHC and give the relevant production cross sections. 	We further discuss the signature of the process $pp\rightarrow Z_D\gamma \rightarrow ll\gamma$ and the SM backgrounds. After simulating the signal as well as the relevant backgrounds, and applying suitable kinematic cuts, the statistical significance can reach 3.4 at the 14TeV LHC with an integrated luminosity of 300 $fb^{-1}$. The numerical results show that the signal of $Z_D$ might be detected at the 14TeV LHC and future collider.

\section*{Acknowledgement}

\noindent
This work was
supported in part by the National Natural Science Foundation of
China under Grants No.11275088 and 11545012, and  the Natural Science Foundation of the Liaoning Scientific Committee under
No. 2014020151.

\end{document}